\documentclass[aps,ajp,twocolumn,longbibliography,showpacs,superscriptaddress]{revtex4-1}
\usepackage{amsmath}
\usepackage[dvipdfm]{graphicx}
\usepackage{amsfonts}
\usepackage{xcolor}
\usepackage{placeins}
\usepackage[bottom]{footmisc}
\usepackage{appendix}
\usepackage{longtable}

\begin{document}

\title{Free Electron Theory for Thin Metal Films}

\author{ Philip B. Allen }
\email{philip.allen@stonybrook.edu}
\affiliation{ Department of Physics and Astronomy,
              Stony Brook University, 
              Stony Brook, New York 11794-3800, USA }

\date{\today}

\begin{abstract}

Quantum free electrons, {\it i.e.} plane waves, with wavevector $\vec{k}$, and
occupancy constrained by the Pauli exclusion principle,
are explained in all solid state physics texts.  Although overly simplified, free-electron theory
works surprisingly well for many properties of simple metals.
For bulk materials, it is assumed that the sample is effectively infinite and that
surfaces are irrelevant.  Over the past 30 years, experiments that visualize
surfaces and enable the study of 2-d materials have revolutionized solid state physics, 
stimulating new experiment, theory, and applications.
Modified free electron models, adapted to films, have enabled modeling of 
electronic properties of films.  This paper analyzes three such models: 
periodic boundary
conditions, 
hard-wall boundary conditions, and 
soft-wall (SW) boundary conditions,
in order of increasing realism.  The SW case is illustrated 
for an aluminum film consisting of six atomic layers, comparing
 SW free-electron theory with a scanning tunneling spectroscopy experiment.

\end{abstract}

\maketitle

\section{Introduction}

The invention \cite{Binnig1982,Binnig1987} of the scanning tunneling microscope enabled the atomic structure 
of crystalline surfaces to be observed.  For example, this allowed 
the mysterious 7 $\times$ 7 reconstruction of the silicon surface
 to be seen in detail \cite{Binnig1983}, and transformed condensed matter physics.  
 Growth and study of crystalline thin films became a new focus of the field. 
 Since this discovery, there have been remarkable improvements in film growth 
(pedagogical discussions can be found in references \cite{Kolbas1984,Zhang1998,Milun2002,Atkinson2008}), 
eventually leading to the discovery of graphene and other two-dimensional (2-d) materials \cite{Novoselov2004}.

This paper focuses on simple crystalline metallic films 
with a well-defined number $N_{\rm L}$ of molecular layers.  
The term ``film'' will be used to mean a thin slab ($N_L$ typically 25 or less).
Each layer is a 2-d crystal, essentially
the same in structure as the corresponding bulk crystal.  Sometimes crystals 
(bulk or slabs, including films) have interesting surface reconstructions, but these will not be considered here.  
For three-dimensional (3-d) crystalline metals, electron excitations are often well described by
the free-electron approximation.  This paper clarifies the corresponding free-electron description of 
thin film electronic excitations, showing how it relates to, and is different from, the
free-electron description of bulk electrons.
 


This paper is organized as follows.  Section \ref{sec:metals} reviews electrons in crystals in the 
single-particle Bloch-state description.
Section \ref{sec:free} discusses nearly-free electrons, resulting in ``electron gases'' in 3-d and 2-d.
Section \ref{sec:cs} discusses 2-d crystals and slabs or films, and introduces the example 
of a 6-layer (111) aluminum (Al) film.
Section  \ref{sec:fes} explains why boundary conditions are needed to describe free electrons in films,
and gives three approximations that enable different versions of the modified free-electron theory, in order of
increasing complexity and accuracy.
Section \ref{sec:sts} discusses 
scanning tunneling spectroscopy, a powerful way of studying electrons in films.  
The measured spectrum of the 6-layer (111) Al film is compared with a computation using
free-electron theory with soft-wall (SW) boundary conditions.
An appendix explains some details needed for use of free-electron theory of films.

\section{Electrons in Crystals}
\label{sec:metals}

Soon after the electron was discovered, Drude \cite{Drude1900} proposed
a theory of electrical conduction in metals using a classical description of
electrons.  Then crystallography was invented, improving understanding of atoms and solids, which allowed
quantum theory to evolve.
Sommerfeld \cite{Sommerfeld1928} gave a quantum version of Drude's model, appropriate for crystalline metals.
Bloch \cite{Bloch1928} immediately enlarged it.  The underlying justification for many of these ideas is
Landau's Fermi-liquid-theory \cite{Landau1959}, which hypothesizes that low-lying electronic 
excitations of simple forms of crystalline matter have a 1-to-1 correspondence to single-particle 
excitations of non-interacting electrons.  

Bloch's theorem 
says that 1-electron states $\psi_{\vec{k}}(\vec{r})$ of electrons in a crystal can be chosen 
as eigenstates of translation, 
with eigenvalues $\exp(i\vec{k}\cdot\vec{\ell})$ and wavevector $\vec{k}$.
The translations $\vec{\ell}=\ell_1\vec{a}+\ell_2\vec{b}+\ell_3\vec{c}$ are integer
multiples of the primitive translations $\vec{a}, \vec{b},\vec{c}$.
These states are accurately computed by versions of the density-functional theory (DFT),
and  describe ground state energies and low-lying excitations of many crystals.  However, 
there are also ``strongly-correlated'' materials where this description is insufficient.  
 Alternatives to Fermi liquid models are widely discussed, but
the subject still is not deeply understood. This paper discusses only ``weakly-correlated'' materials,
where Fermi liquid theory and single-particle Bloch states work well.


\section{nearly free electrons} 
\label{sec:free}

\subsection{3-d Crystals}

The ``single-particle approximation'' uses the electronic Hamiltonian
\begin{equation}
\mathcal{H}_{\rm el} = \sum_i p_i^2/2m + \sum_{i,\vec{\ell},a}V_a(\vec{r}_i-\vec{R}_{\vec{\ell} a})
\label{eq:}
\end{equation}
where $p_i=|\vec{p}_i|$ is the momentum of electron $i$ of mass $m$ and position $\vec{r}_i$,  
$\vec{\ell}$ labels unit cells, $a$ lists atoms in the cell, and $\vec{R}_{\vec{\ell} a}$ 
is the equilibrium position of the atom.  There are no explicit electron-electron Coulomb repulsions -- the
repulsion is assumed to be absorbed in the effective electron-ion
interactions $V_a(\vec{r}_i-\vec{R}_{\vec{\ell} a})$.  The ``free-electron approximation'' assumes small 
electron-ion interactions $V_a$, and omits them, keeping only the kinetic energy. 
Then the single particle wavefunctions and energies are
\begin{equation}
\psi_{\vec{k}}(\vec{r}) = \sqrt{\frac{1}{\Omega}} e^{ i\vec{k}\cdot\vec{r} }; \ \ \ \epsilon_k=\frac{\hbar^2 k^2}{2m},
\label{eq:psife}
\end{equation}
where $\Omega$ is the volume of the sample.  This simplification describes many properties of
``simple metals'' like Na, Mg, Al, and Pb, with valence electrons derived from atomic
$s$ and $p$ states.  Alkali metals have Fermi surfaces very close to the 
free-electron sphere.  Even in metals like Al with more than one valence electron per atom,
and distorted Fermi spheres, most low-lying excitations are not much affected
by the details contained in the potential $V_a$, and the free-electron approximation works well for many properties. 

\subsection{2-d Free Electron or Hole Gases}

Dilute 2-d nearly-free-electron gases can be made by confining electrons or holes near the surface of a
semiconducting material.  Confinement requires an external potential, provided usually by
an interface \cite{Ando1982,Sandomirskii1989} or gate electrodes
\cite{Gabay2013,Ambacher1999,Ohtomo2004}.  Examples can be found in recent papers \cite{Dill2025,Khan2025}.  
The current paper discusses nearly-free-electron metals where confinement in the $z$-direction results
simply from the thinness of the sample, with charge density essentially the same as in a bulk sample.

\section{2-d Crystals and Films}
\label{sec:cs}

\subsection{Mono- and Bi-layer 2-d Crystals}

A 2-d crystal has translations $\vec{\ell}=\ell_1\vec{s}_1+\ell_2\vec{s}_2$.  These are usually
a subset of the translations of a parent 3-d crystal. 
Graphene was the first monolayer 2-d crystal produced \cite{Novoselov2004}, made by stripping a single layer of 
hexagonally arranged carbon atoms from the parent crystal of graphite.  Graphite is a 3-d crystal
consisting of many layers of graphene, stacked
periodically (in the $\hat{z}$-direction, perpendicular to the graphene plane.)  
Layers of graphite couple by weak van der Waals interactions, making removal of a single layer
easy.   Bilayer graphene \cite{Novoselov2004} and transition metal dichalcogenides (like 
WSe$_2$) \cite{Manzeli2017} are other examples of 2-d crystals.
Fabrication methods and properties of these materials are now much studied.  
Few-layer metallic 2-d crystals are harder to make \cite{Wang2020,Yu2023}.   
Simple-metal mono- and bi-layer 2-d crystals are now close to being successfully synthesized \cite{Zhao2025}.
This paper, however, focuses on the situation of a slab of a crystalline material.

\subsection{Crystalline Slabs and Films}

A crystalline slab is a slice of a crystal, with dimensions $N_1\vec{s}_1$ ,$N_2\vec{s}_2$, and $N_3\vec{s}_3$.
The in-plane vectors $\vec{s}_1$ and $\vec{s}_2$ are the two smallest combinations of the primitive 3-d
vectors $\vec{a}, \vec{b}, \vec{c}$ that lie in the plane of the slice, and the third translation $\vec{s}_3$ is chosen
so that any crystal translation can be written as $\Sigma_i \ell_i\vec{s}_i$.  This paper uses the words
``horizontal'' to mean in-plane,``vertical '' to mean perpendicular to the plane. 
The layer separation is $h=\hat{z}\cdot\vec{s}_3$, and $H=N_L h$ is the thickness of the slab. 
For films of simple metals with $N_L\gtrsim 25$, the interior (more than a few layers away 
from either surface) is mostly well described by the ordinary free-electron model.  But for thinner
films, or for properties near a surface of a slab, the free-electron picture requires modification.

\subsection{6-layer Aluminum (111) Film} \label{sec:Al111}
\label{sec:Al}

\par
\begin{figure}
\includegraphics[angle=0,width=0.44\textwidth]{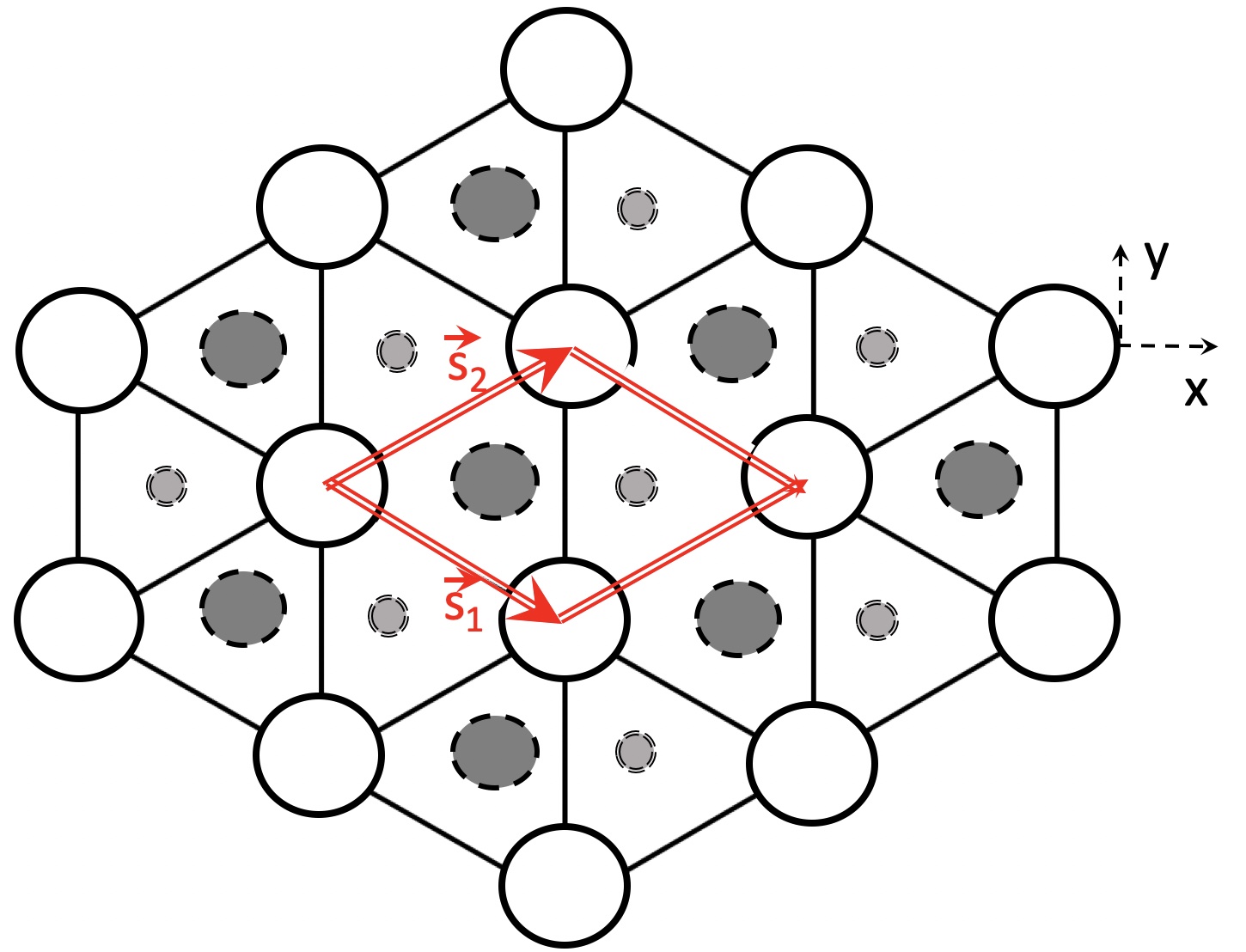}
\caption{\label{fig:fcc111} Three (111) layers of an fcc crystal  seen from above.  
The nearest neighbor distance is $a/\sqrt{2}=|\vec{s}_1|=|\vec{s}_2|=2.86 \ \AA$ (for Al).
Layer 2 (medium-size circles) and 3 (small circles)  lie at distances $h=a/\sqrt{3}$
and $2h$ below layer 1.}
\end{figure}
\par

Many metallic films have been studied experimentally, including both {\it{s}} and {\it{p}}-metals, 
and {\it{d}} and {\it{f}}-band metals.
This paper uses the example of aluminum, and specifically an Al film of six layers ($N_L$=6), 
as has been experimentally studied by Wu {\it et al.} \cite{Wu2015} and Cheng {\it et al.} \cite{Cheng2016,Cheng2018}.
Such a film is an idealized piece of a single face-centered-cubic (fcc) Al crystal, sliced parallel
to the (111) plane, and shown in Fig. \ref{fig:fcc111}.
Each layer of the film is a plane of triangular-close-packed atoms, arranged
in rhombic cells at sites $\vec{\ell}=\ell_1\vec{s}_1+\ell_2\vec{s}_2$, where
\begin{equation}
\vec{s}_1=\frac{a}{\sqrt{2}}\left(\frac{\sqrt{3}}{2}\hat{x}-\frac{1}{2}\hat{y}\right); \  \  \ 
\vec{s}_2=\frac{a}{\sqrt{2}}\left(\frac{\sqrt{3}}{2}\hat{x}+\frac{1}{2}\hat{y}\right).
\label{eq:abprime}
\end{equation}
In fcc crystals, the $x,y,z$ directions are normally defined to
point along the axes of the cube (with volume $a^3$).  For films it is sensible
 to have $\hat{z}$ pointing perpendicular to the film plane, 
 and $\hat{x}$ and $\hat{y}$ lying in the plane of the film, as shown in Fig. \ref{fig:fcc111}.
The Al nearest neighbor distance is $a/\sqrt{2}=|\vec{s}_1|=|\vec{s}_2|=2.86 \ \AA$, where
$a=4.05 \ \AA$ is the fcc lattice constant.  
The layer spacing (in the $\hat{z}$ direction) is $h=a/\sqrt{3}=2.34 \ \AA$.
The second and third layers are identical to the first, except lower by $h$ and $2h$, 
and shifted (in plane) by 1/3 and 2/3 of the diagonal vector ($\vec{s}_1+\vec{s}_2$) 
of the rhombic cell.  The fourth layer is exactly below the first layer by $3h$.  
In actual crystals, the layer spacing will relax slightly near the surface, but this is hard to 
measure or compute and will be ignored here for simplicity.

\section{Modifications of Free electron model for films}
\label{sec:fes}

Let the in-plane boundaries of the film be at $N_1\vec{s}_1$ and
$N_2\vec{s}_2$, with sample area $\mathcal{A}=N_1 N_2 |\vec{s}_1 \times \vec{s}_2|$.
Let $N_1$ and $N_2$ be large enough ({\it i.e.} $200$ or more)
so that the influence of surfaces perpendicular to the xy plane can be ignored. 
In-plane, the electrons are free and, using periodic in-plane boundary conditions, 
the electron wavefunctions can be written as
$\psi = \mathcal{A}^{-1/2} \exp(i\vec{k}\cdot\vec{r}_\parallel) u(z)$.  
In-plane wavevectors $\vec{k}$ are well approximated as a continuum.
The out-of-plane component $u(z)$ is still considered to be ``free'' ({\it{i.e.}}
not affected by the crystal potential $V_n$), but is more complicated.
Because the thickness is of nanometer scale ($N_3 < 25$), the perpendicular wavevectors
are discrete (labeled $k_n$ or $k_{\perp n}$), and propagation in the $z$-direction is limited.

Three simple boundary conditions are now discussed, which can be used to 
implement the free-electron approximation.  
In order of increasing complexity, they are: 
{\bf {A}}: periodic (PBC), {\bf {B}}: hard-wall (HW), and {\bf{C}}: soft-wall (SW).  
Only the SW case allows the interior of the film to be influenced by vacuum or other
media above and below the film.  The other two forbid such influence.

\subsection{Periodic Boundary Conditions}

Adjacent layers of a slab are separated by a vertical distance $h$.  The positions of slab surfaces are 
chosen to be at distance $h/2$ below the bottom layer of atoms, and $h/2$ above 
the top layer.  The $N_L$-layer slab then has surfaces a distance $H=N_L h$ apart.  
Periodic boundary conditions (PBC) force slab wavefunctions to have
$\psi(\vec{r}_\parallel,z)= \psi(\vec{r}_\parallel,z+H)$.  This is the simplest, most used, but least
realistic, of the three boundary conditions described in this paper.  The resulting free-electron states are
\begin{eqnarray}
&&\psi_{\vec{k}n}=\frac{1}{\Omega}e^{i\vec{k}_\parallel\cdot\vec{r}_\parallel+ik_{\perp n} z} \nonumber \\
&&\epsilon_{kn}=\frac{\hbar^2}{2m}[k_\parallel^2+ k_{\perp n}^2]
\equiv\frac{\hbar^2}{2m}k_\parallel^2 + \epsilon_n^{\rm PBC} \nonumber \\
&&\vec{k}_\parallel=n_1\vec{g}_1/N_1 + n_2\vec{g}_2/N_2; \ \ k_{\perp n}=2\pi n/H \nonumber \\
\label{eq:kpbc}
\end{eqnarray}
where $\vec{g}_1$ and $\vec{g}_2$ are the 2-d reciprocal lattice vectors,
and $n_1,n_2,\rm{and} \ n$ are integers (including 0 and negative values).

\begin{figure}
\includegraphics[angle=0,width=0.41\textwidth]{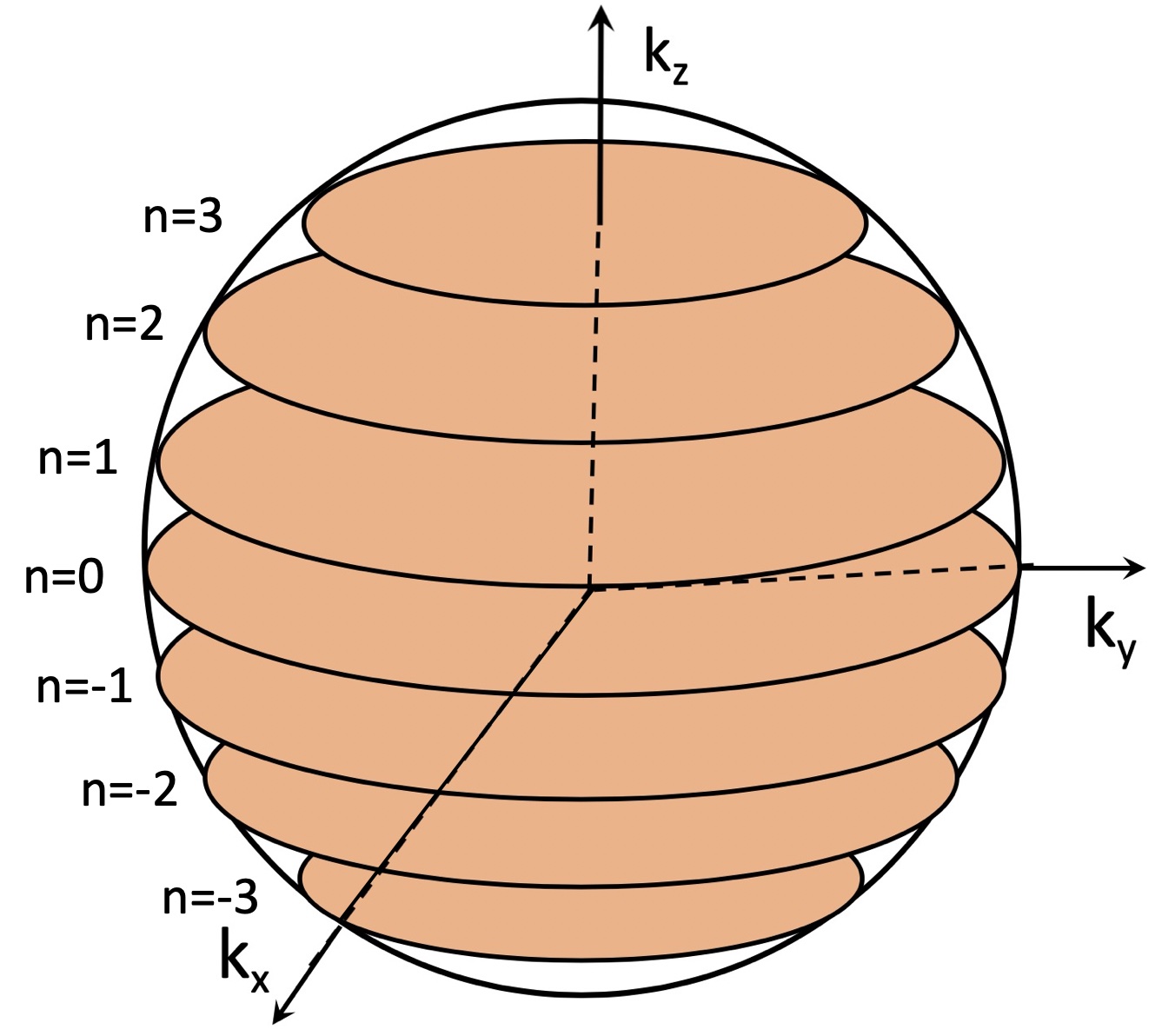}
\caption{\label{fig:evsk2dAlPBC} The edges of the brown disks are the free-electron  
``Fermi surface'' of the 6-layer Al(111) film, as given by periodic boundary conditions. 
Occupied states lie in circular disks stacked at values of $k_z$ equal to $n(2\pi/H)$.   
In the bulk limit ($N_{\rm L}\rightarrow \infty$), the occupied circles are closely spaced, essentially 
filling the free electron Fermi sphere.} 
\end{figure}

The occupied states are illustrated in Fig. \ref{fig:evsk2dAlPBC}, for the case of a 6-layer ($N_L=6$) 
Al film.  The circular disks contain the occupied states of each disk $n$, with separate 
radii $k_{Fn}$. 
The Fermi energy $\epsilon_{F,N_L}=(\hbar k_{Fn})^2/2m + \epsilon_n^{\rm PBC}$ 
is the same in each disk.
The number of occupied disks ($2n_{\rm max}+1$ for PBC's) and the film's Fermi energy $\epsilon_{F,N_L}$, 
require computation.  The value of $\epsilon_{F,N_L}$ is not quite the same as 
$\epsilon_F$ of the bulk metal; it varies with $N_L$.  Details are in the appendix.
For the 6-layer film of Al, $n_{\rm max}=3$, the Fermi energy is $0.6\%$ higher than the value of
11.63 eV for bulk Al, and the Fermi wavevector $k_{F0}$ of the $n=0$ disk is $0.3\%$ larger than the bulk value
$k_F=1.75 \times 10^8$ cm$^{-1}$.

\par
\par

\subsection{Hard Wall Boundary Conditions}

Hard walls (HW) force $\psi$ to vanish at the surfaces.
It is convenient to choose the locations of the surfaces to be $z=0$ and $z=H$.
The hard-wall states are
\begin{eqnarray}
&&\psi_{\vec{k}_\parallel ,n}(\vec{r})=\sqrt{\frac{2}{\Omega}}e^{i\vec{k}_\parallel \cdot \vec{r}_\parallel}\sin(k_n z); 
\nonumber \\ 
&&k_n=\frac{n\pi}{H}; \ \  \epsilon_{kn}=\frac{(\hbar k_\parallel)^2}{2m} + \epsilon_n^{\rm HW} \nonumber \\
&&  \epsilon_n^{\rm HW} = \frac{(\hbar k_n)^2}{2m}
\label{eq:psiHW}
\end{eqnarray}
The energy is the same as in Eq. \ref{eq:kpbc}, except that the $z$-propagating wavevector $k_{\perp n}$ is
replaced by a similar but non-propagating wavevector $k_n$, with $n>0$ now a positive integer.

The occupied Fermi disks are shown in Fig. \ref{fig:evsk2dAlHW} for 6-layer Al.
There are $n_{\rm occ}=8$ partially filled disks, as derived in the appendix.  
The Fermi energy, $\epsilon_F^{\rm HW}$=12.45 eV, is 6.9\% higher than in bulk Al.  

\par
\begin{figure}
\includegraphics[angle=0,width=0.42\textwidth]{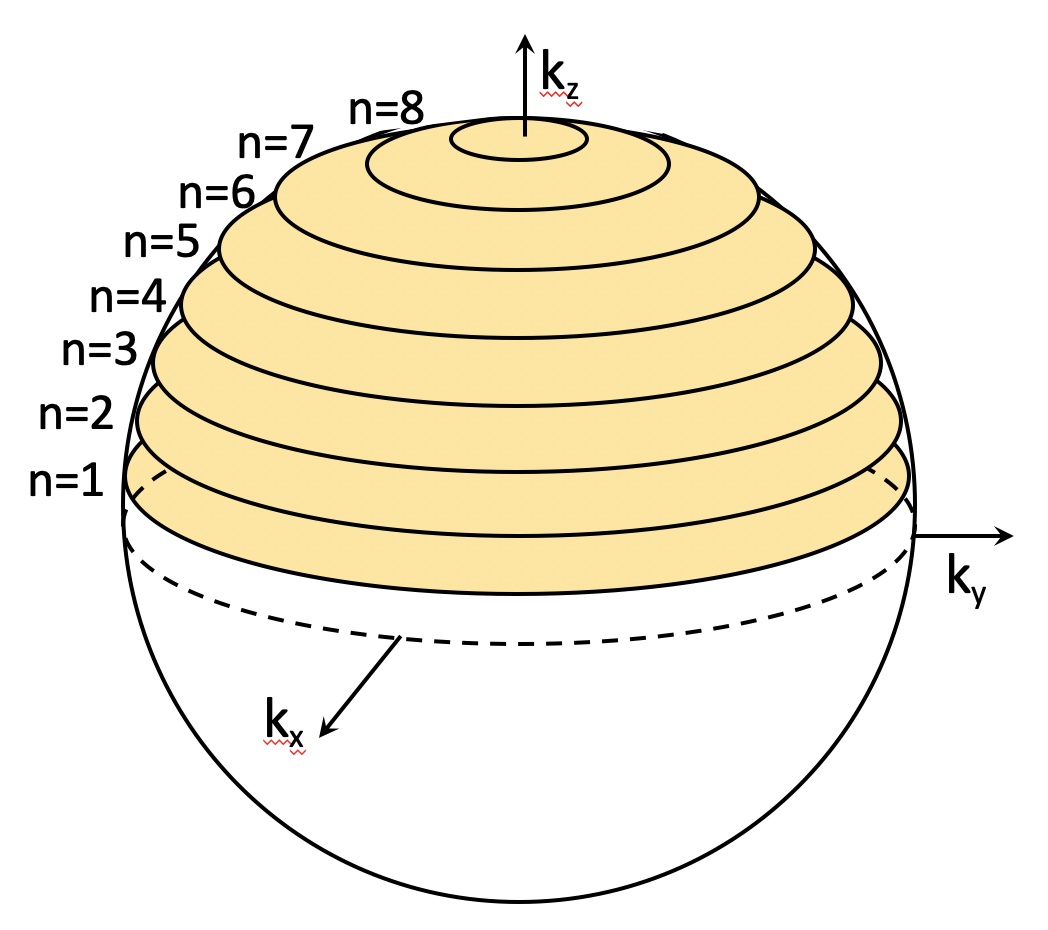}
\caption{\label{fig:evsk2dAlHW} The edges of the yellow disks are the free-electron  
``Fermi surface'' of the 6-layer Al(111) film, as given by hard-wall boundary conditions. 
Occupied states lie in circular disks stacked at values of $k_z$ equal to $n(\pi/H)$.  
A similar figure is in Ref. \onlinecite{Czoschke2005}.}
\end{figure}
\par

\subsection{Soft Wall Boundary Conditions}

The soft wall (SW) model confines low energy electrons by use of a $z$-dependent 
(``quantum well'') potential
that is constant (set here as the 0 of energy) inside the film, and a higher constant potential $E_{\rm vac}$,
the vacuum energy, outside the film.  The HW model uses $E_{\rm vac}=\infty$.  
If one (or both) outside regions have a substrate that is not vacuum,
a less simple SW model can be used \cite{Milun2002}.
It is now convenient to define the location of the surfaces to be at $z=\pm H/2$, so that the 
potential is
\begin{equation}
  V_{\rm SW}(z)=\begin{cases}
    0 & \text{if $|z|<H/2$},\\
    E_{\rm vac} & \text{otherwise},
  \end{cases}
\label{eq:VSW}
\end{equation}
The interior of the film is no longer disconnected from the vacuum outside.   
These boundary conditions are more realistic than the HW model, and hardly change the 
low-lying electron states within the film.

Excited electrons ($\epsilon>\epsilon_F$) cannot escape to the vacuum unless their 
energy exceeds $E_{\rm vac}$.   The difference, $E_{\rm vac}-\epsilon_F$
is the ``work function'' $\Phi$, the minimum photon energy $\hbar\omega$
which can eject an electron from the material.
The value of $\Phi$ depends on the nature of the surface; the work function of Al with a clean
(111) surface is $\Phi(111)$=4.26 eV \cite{Eastment1973}. 
In a thin film, finite thickness can slightly alter the work function \cite{Boettger1996,Kim2010}, but this
is ignored in the simple model discussed here.

The wavefunction of Eq. \ref{eq:psiHW} is replaced by
\begin{equation}
\psi_{\vec{k}_{\parallel,n}}(\vec{r},z)=\exp(i\vec{k}_\parallel\cdot\vec{r})u_n(z).
\label{eq:psiSW}
\end{equation}
Here $u_n(z)$ is the solution of the Schr\"odinger equation of a particle in a quantum well,
\begin{equation}
\left[ -\frac{\hbar^2}{2m} \left(\frac{\partial}{\partial z}\right)^2+V_{\rm SW}(z)\right] u_n(z) = \epsilon_n^{\rm SW} u_n(z).
\label{eq:uSW0}
\end{equation}
Unlike the HW case, the wavefunction $u_n(z)$ does not vanish outside the film.  
Continuity of $\psi$ at the edges $z=\pm H/2$
forces the wavevector $\vec{k}_\parallel$ to be the same
inside and outside the film.  When $\epsilon_n^{\rm SW} < E_{\rm vac}$, the wavefunction
$u_n$ decays exponentially outside the film; when  $\epsilon_n^{\rm SW} > E_{\rm vac}$, the part of $u_n(z)$ 
outside the film propagates freely.

The energies are
\begin{equation}
\epsilon_{\vec{k}_\parallel n}=\frac{\hbar^2}{2m}k_\parallel^2 + \epsilon_n^{\rm SW}; \ \ 
\epsilon_n^{\rm SW}=\frac{\hbar^2 \alpha_n^2}{2m}.
\label{eq:epsSW}
\end{equation}
The SW states $n$ line up with their corresponding HW states (Eq. \ref{eq:psiHW}).  
The Fermi level is significantly below the vacuum level.  Since the excited states
($\epsilon_{kn}^{\rm SW}>\epsilon_F$) discussed below have $\epsilon_{kn}^{\rm SW}<E_{\rm vac}$, 
we only need to consider confined states.  The wavefunctions are then
\begin{eqnarray}
u_n(z)&=& 
    \begin{cases}
       &\sin( \alpha_n z) \ ({\rm odd})\\
       &\cos ( \alpha_n z) \ ({\rm even})\\
         \end{cases} \ {\rm if} \ |z|\le H/2,  \nonumber \\
u_n(z) &=&
C_n e^{-\beta_n(|z|-H/2)} \ \ \ {\rm if} \ |z| > H/2 .
\label{eq:uSW}
\end{eqnarray}
The decay coefficient is $\beta_n=\sqrt{2mE_{\rm vac}/\hbar^2 - \alpha_n^2}$.

\begin{figure}
\includegraphics[angle=0,width=0.49\textwidth]{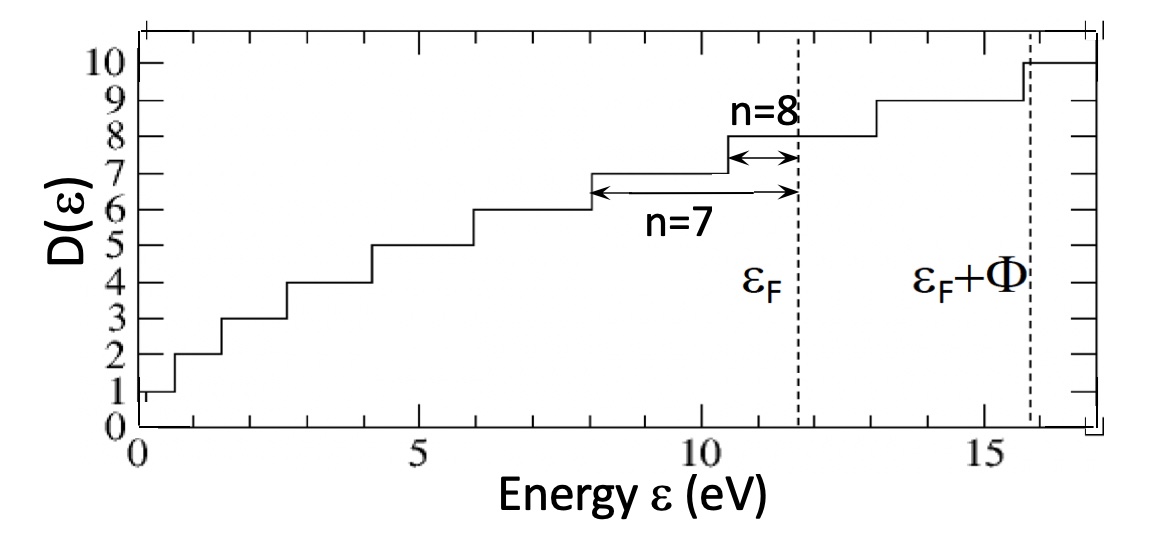}
\caption{\label{fig:SWDOS}  Density of states (DOS, in units of the 2-dimensional DOS 
of Eq. \ref{eq:D2}) {\it vs.} energy for the 6-layer Al film with SW boundary conditions. 
The energy starts at the bottom of the bulk free-electron band ($E=0$) and goes past the bulk Fermi energy 
($\epsilon_F$ = 11.67 eV), up to the vacuum energy ($\epsilon_F +\Phi$, $\Phi$ = 4.26 eV). 
Horizontal arrows clarify which transitions give the peaks in $dI/dV$ discussed in Sec. \ref{sec:sts}
and shown in Fig. \ref{fig:dIdV}(b). 
The calculations use SW energy levels given in Table I.}
\end{figure}

Forcing continuity of $\psi$ and $d\psi/dz$ across the boundaries gives for the even states (odd $n$)
\begin{equation}
\alpha_n \tan(\alpha_n H/2)=\sqrt{2mE_{\rm vac}/\hbar^2 - \alpha_n^2}.
\label{eq:A3}
\end{equation}
and for the odd solutions (even $n$)
\begin{equation}
-\alpha_n \cot(\alpha_n H/2)=\sqrt{2mE_{\rm vac}/\hbar^2 - \alpha_n^2}.
\label{eq:A4}
\end{equation}
These two equations can be solved numerically,
and have a finite number of solutions $\alpha_n > 0$,
ending when $\alpha_n$ exceeds $\sqrt{2mE_{\rm vac}/\hbar^2}$.
These solutions for a square well potential are given in standard
texts, for example, Ref. \cite{Schiff1955} eqs. 9.6 and 9.7.
The wavenumbers $\alpha_n$ are close to the wavevectors $k_n=n\pi/H$ of the hard-wall case; 
soft-wall energies are fairly close to the hard-wall values of Eq. \ref{eq:psiHW}, but lower in energy
because the $z$-range of $\psi$ is increased.  Results are given in Table I for the 6-layer Al film.
The number of occupied
disks ($n_{\rm max}$) is 8 (the same as the HW case), but SW disks are not identical to HW disks shown in
Fig. \ref{fig:evsk2dAlHW}, because the $k_z$-coordinates $\alpha_n$ are
slightly different from the HW $k_n$.  The Fermi energy is $\epsilon_F^{\rm SW}$=11.72 eV,
less than $\epsilon_F^{\rm HW}$=12.45 eV, and accidentally the same  as $\epsilon_F^{\rm PBC}$,
0.6 \% higher than in bulk Al.  Derivations are given in the appendix.
The SW electronic density of states of 6-layer Al is shown in Fig. \ref{fig:SWDOS}.
 
\par
\par
\begin{table}[h!]
\centering
\begin{tabular}{|r| r r| r r| r r|} 
 \hline
   & Hard &Soft & Hard& Soft & Soft& \ \ \ \ \ \ \ \ \  \\
 \hline
 n & $k_n a$  & $\alpha_n a$ & $\epsilon_n^{\rm HW}$ & $\epsilon_n^{\rm SW}$ & $\beta_n a$ 
 & $f_n^{\rm out} \ \ \ \ $ \\ [0.5ex] 
 \hline
 1 & 0.91 & 0.85 &0.19 & 0.17 & 8.23 & 3.45$\times 10^{-4}$\\ 
 2 & 1.81 & 1.69 & 0.76 & 0.66 & 8.10 & 1.40$\times 10^{-3}$\\
 3 & 2.72 & 2.54 & 1.72 & 1.50 & 7.88 & 3.24$\times 10^{-3}$\\
 4 & 3.62 & 3.38 & 3.05 & 2.16 & 7.55 & 5.99$\times 10^{-3}$\\
 5 & 4.53 & 4.23 & 4.77 & 4.14 & 7.12 & 8.52$\times 10^{-3}$\\ 
 6 & 5.44 & 5.06 & 6.87 & 5.95 & 6.55 & 1.54$\times 10^{-2}$\\
 7 & 6.35 & 5.89 & 9.35 & 8.05  & 5.82 & 2.35$\times 10^{-2}$\\
 8 & 7.26 & 6.71 &12.21 &10.44 & 4.85 & 3.63$\times 10^{-2}$\\
 9 & 8.16 & 7.51 &15.45 &13.07 & 3.49 & 6.20$\times 10^{-2}$\\
 10& 9.07 & 8.23 &19.08&15.70 & 0.92 & 2.35$\times 10^{-1}$\\ 
 \hline
\end{tabular}
\caption{For the 6-layer Al(111) film, these are the perpendicular wavenumbers ($k_n,\alpha_n$)
and energies (in eV) for the first 10 states under HW and SW boundary conditions.
Formulas for $k_n$ and $\epsilon_n^{\rm HW}$ are given in Eq. \ref{eq:psiHW}.
The last two columns give the decay constants $\beta_n$ and the
probability $f_n^{\rm out}$ (Eq. \ref{eq:fout})
for the soft-wall wavefunction $u_n$ of Eq. \ref{eq:uSW} to lie outside
the film with $|z|>H/2$.
The free-electron Fermi energy of bulk Al is 11.67 eV, and the vacuum level ($\epsilon_F+\Phi$) is 15.93 eV. }
\label{table:1}
\end{table}

``Confined states'' ($\epsilon<E_{\rm vac}$) are well-confined if $\epsilon$ is significantly below $E_{\rm vac}$; 
higher energy confined states have a significant fraction of their probability 
distribution outside the film.  This probability, designated as $f_n^{\rm out}$, is given by
\begin{equation}
f_n^{\rm out}=\int_{|z|>H/2} dz |u_n(z)|^2/\int_{-\infty}^\infty dz |u_n(z)|^2.
\label{eq:fout}
\end{equation}
Values of $f_n^{\rm out}$ are shown in table I for the quantum-well-confined states of 6-layer Al (111).
As can be expected, the higher n, the larger the probability for the electron to be outside the slab. Even at n=8
(just below the Fermi energy), an electron has almost 4\% of its probability outside the film.


\section{Scanning Tunneling Spectroscopy} 
\label{sec:sts}

\par
\begin{figure}[b]
\includegraphics[angle=0,width=0.50\textwidth]{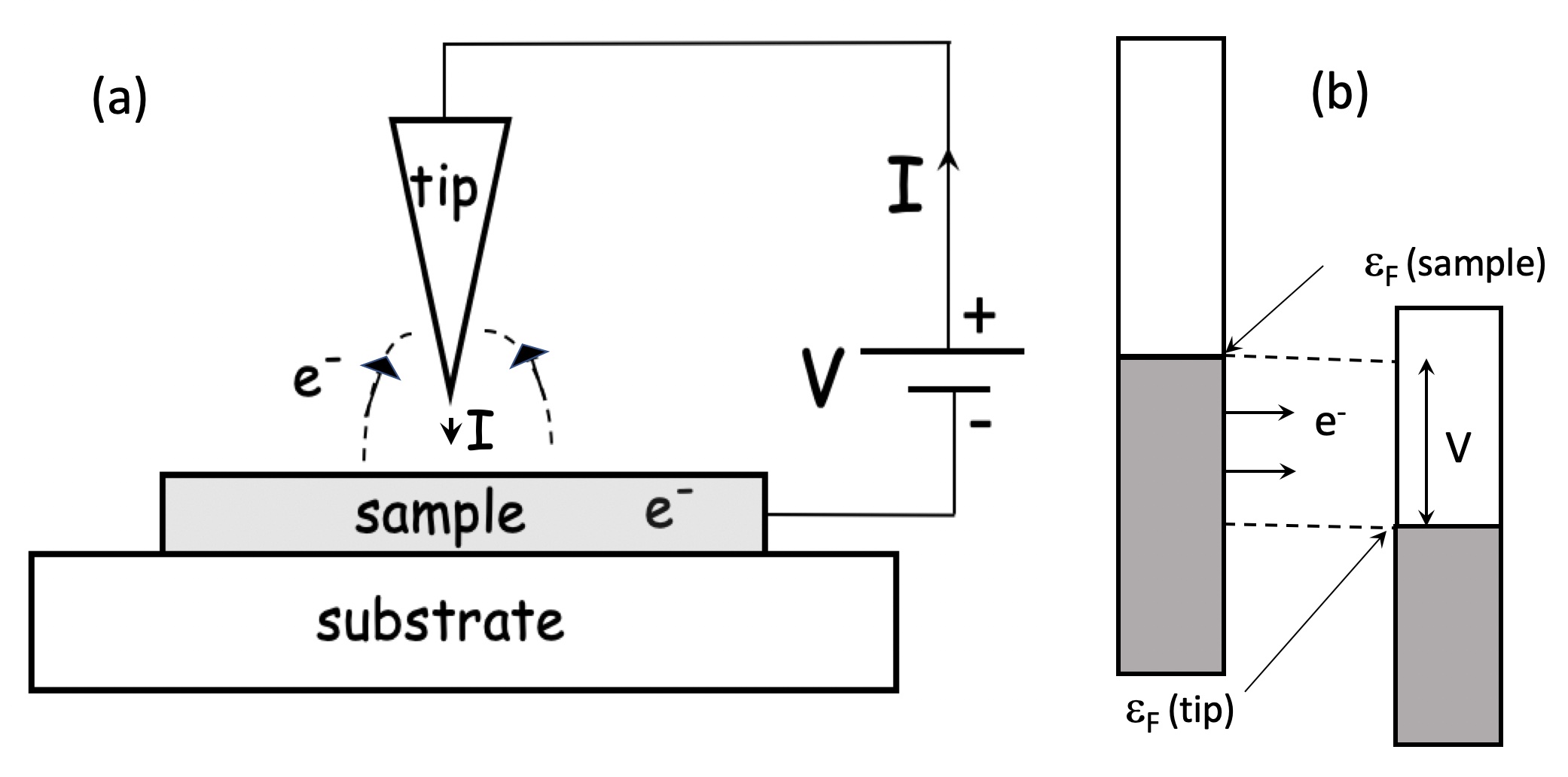}
\caption{\label{fig:STM} 
 (a) Schematic of the scanning tunneling microscope (STM).  The space outside sample, subtrate, and tip
 is vacuum.  (b) Energy states of sample and tip under applied voltage $V$.
 When $V$ is positive, a few electrons will tunnel
out of occupied states of the sample into empty states of the tunneling tip.
Because $V$ is less than the work function $\Phi$ of the sample, the tunneling
electrons do not propagate freely in the vacuum, but decay exponentially away from the sample.}
\end{figure}
\par

\begin{figure}[b]
\includegraphics[angle=0,width=0.50\textwidth]{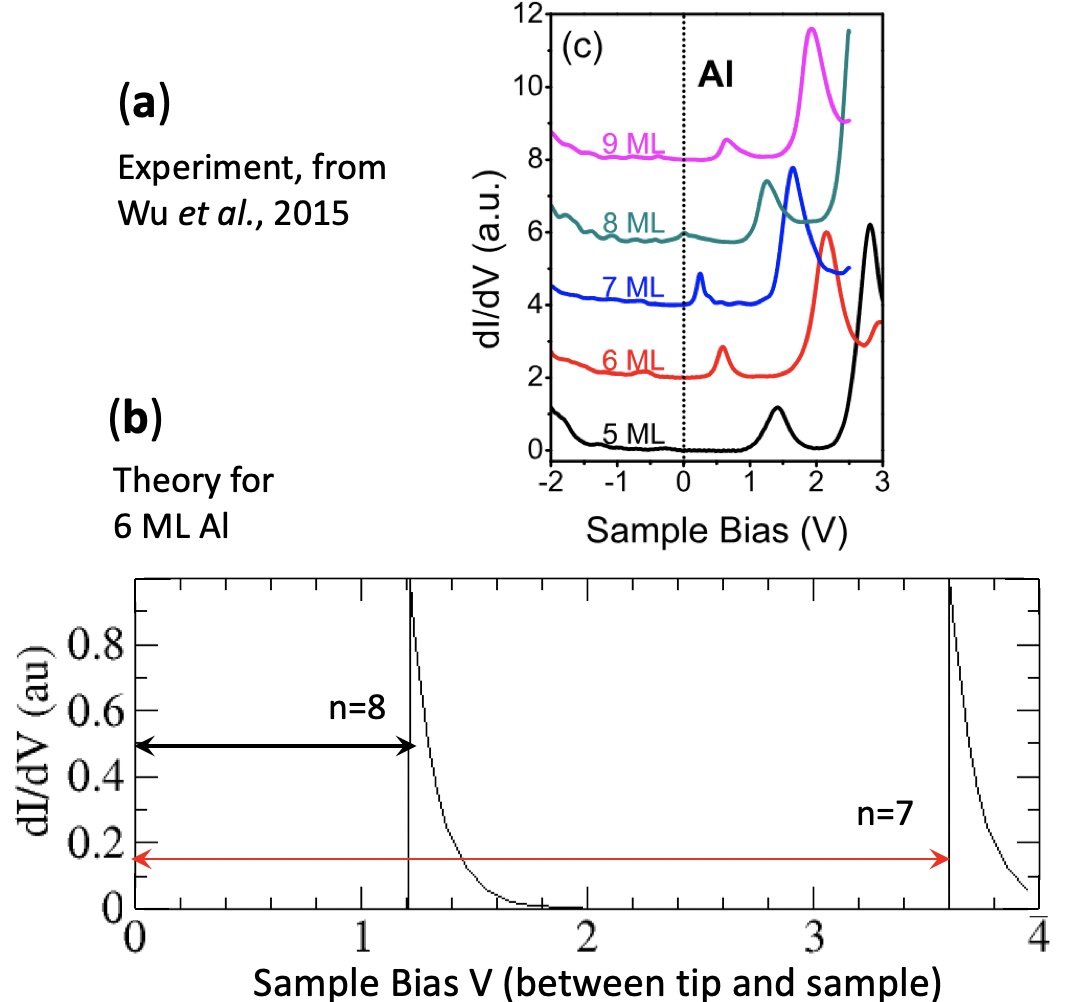}
\caption{\label{fig:dIdV} (a)  Experimental STM spectroscopy results for five different (111) films of Al.  
The peaks (sometimes called ``quantum well states'') shift as the number $N_{\rm L}$ of 
monolayers (ML) changes.  From Fig 4(c) of Ref. \onlinecite{Wu2015}.
(b) Predicted scanning tunneling spectrum $dI/dV$ (in arbitrary units) from Eq. \ref{eq:dIdV}
for 6 ML Al, using an empirical parameter $\gamma$ = 8.3 eV$^{-1}$.  
Horizontal arrows clarify which disk states shown in Fig \ref{fig:SWDOS}
give the peaks corresponding with those measured in the 6 ML case in (a).}
\end{figure}

Figure \ref{fig:STM} illustrates a scanning-tunneling microscope (STM).  When close to the sample, the STM 
tip couples to the spatial part of occupied electron wavefunctions that lie outside the film.  
Thus PBC and HW boundary conditions cannot be used to model STM spectra. 
This section uses SW free electron states and energies (Eqs. \ref{eq:epsSW}-\ref{eq:A4}) to model the STM 
$dI/dV$ spectra of a 6-layer Al (111) film (theory and experiment are shown in Fig. \ref{fig:dIdV}).

Formulas for the tunneling between source and tip are complicated; explanations are in Refs.  
\onlinecite{Tersoff1983} and \onlinecite{Chen2021}.
The tunneling current requires overlap between occupied electron states $k_s=(\vec{k}_{s\parallel},n_s)$ 
of the sample, and empty states $j_t$ of the tip that receive tunneled electrons.  There is a potential energy
$V({\rm tip,sample})$ and a corresponding matrix element $M_{k_s,j_t}=<k_s|V({\rm tip,sample})|j_t>$.
Then summing over all available states gives the  
tunneling current $I$ as a function of the tip-sample voltage $V$ (and potential energy difference
$-eV$).  At temperature $T=0$, the formula is
\begin{equation}
I(V) \propto \sum_{k_s}^{\rm occ}   \sum_{j_t}^{\rm empty}
\delta (\epsilon_{k_s} +eV - \epsilon_{j_t}) |M_{k_s, j_t}|^2.
\label{eq:IV1}
\end{equation}
The Fermi energies $\epsilon_{Fs}$ of the sample and $\epsilon_{Ft}$ of the tip are the same 
when the applied voltage $V$ is zero, and no current flows.  The sign of $V$ is chosen here so that
positive $V$ lowers the Fermi energy of the tip relative to that of the sample, allowing electrons to
tunnel from sample to tip.  The current direction is then from tip to sample, defined as positive current $I$.

A realistic treatment of the tunneling matrix element $M_{k_s, j_t}$ is a 
huge challenge.  The potential $V({\rm tip,sample})$ 
requires a detailed model of the tip, and $|j_t>$ requires DFT calculations for the complicated tip geometry.
$V({\rm tip,sample})$ decays rapidly with depth into the sample, so when computing the matrix elements,
one can use only the parts of states $|k_s>$ decaying exponentially above the surface.  A simplified matrix element is
\begin{equation}
 |M_{k_s, j_t}|^2 \propto e^{-(k_{s\parallel}/\Delta k_\parallel)^2}\equiv e^{-\gamma\epsilon_\parallel},
  \label{eq:Msq}
\end{equation}
where $\epsilon_\parallel=\hbar^2 k_{s\parallel}^2/2m$.
This is based on models (\cite{Boettger1996},\cite{Kim2010},\cite{Baratoff1984}) indicating
that states that succeed in tunneling through the
vacuum region have parallel wavevectors $\vec{k}_\parallel$ confined to a small 
region $|\vec{k}_\parallel | \approx\Delta k_\parallel$ near $\vec{k}_\parallel=0$.  
This generates a Gaussian width $\gamma$, which is treated as an adjustable parameter.
Details of the tip states are ignored.  It is assumed that for each tunneling electron,
the tip states tunneled into are distributed evenly in energy, and similar to each other
even as the energy of the extracted electron varies.  Then the current can be written 
\begin{eqnarray}
I(V) &\propto&\sum_{n=n_{\rm min}}^{n_{\rm max}} \sum_{\vec{k}_\parallel} 
\Theta[\epsilon_{Fs}-(\epsilon_{\vec{k}_\parallel} + \epsilon_n)] \nonumber \\
&\times& \Theta[(\epsilon_{\vec{k}_\parallel}+\epsilon_n)-(\epsilon_{Fs}-eV)]
e^{-\gamma\epsilon_\parallel}.
\label{eq:IV2}
\end{eqnarray}
The step functions $\Theta$ enforce the rule that, since tunneling electrons maintain constant energies,
only sample electrons not more than $eV$ below $\epsilon_{Fs}$ of the sample
can tunnel into the empty states above $\epsilon_{Ft}$ of the tip;
$n_{\rm min}$ and $n_{\rm max}$ are the smallest and the largest sample disk indices $n$ that satisfy this.
The sum over $\vec{k}_\parallel$ can be converted into an integral over $\epsilon=\epsilon_{\vec{k}_\parallel}$:
\begin{equation}
I(V) \propto \sum_{n=n_{\rm min}}^{n_{\rm max}} \int_{\epsilon_{Fs}-\epsilon_n -eV}^{\epsilon_{Fs}-\epsilon_n} 
d\epsilon \mathcal{D}_2(\epsilon) e^{-\gamma \epsilon}
 \label{eq:IV3}
\end{equation}
where $\gamma$ is an adjustable parameter.
Here $\mathcal{D}_2(\epsilon)=\mathcal{D} _2$ is the density of states of a  2-d  free-electron gas
\begin{equation}
\mathcal{D}_2(\epsilon)=\sum_{\vec{k}_\parallel} \delta(\epsilon-\frac{(\hbar k_\parallel)^2}{2m})=
\frac{2\pi \rho_{2d}}{\epsilon_F}=\mathcal{D}_2.
\label{eq:D2}
\end{equation}
Here $\rho_{2d}$ is the electron density of a single disk (by definition the same for all disks).  
The free-electron density of states of an $N_{\rm L}$-layer sample is then
\begin{equation}
\mathcal{D}_{N_{\rm L}}(\epsilon) = \mathcal{D}_2 \sum_{n=1}^\infty \theta(\epsilon-\epsilon_n).
\label{eq:DNL}
\end{equation}
This is plotted in Fig. \ref{fig:SWDOS}, using the energies $\epsilon_n=\epsilon_n^{\rm SW}$
from table I.

\par
\par

The sum in Eq. \ref{eq:IV3} simplifies further when 
the voltage derivative $dI/dV$ is taken:
\begin{equation}
\frac{dI}{dV} \propto \sum_{n=n_{\rm min}}^{n_{\rm max}} e^{-\gamma(eV+\epsilon_{Fs}-\epsilon_n)} 
\label{eq:dIdV}
\end{equation}
This is plotted in Fig. \ref{fig:dIdV}(b), and can be compared with the 6 monolayers spectrum in 
Fig. \ref{fig:dIdV}(a), which shows two ``quantum well states" appearing when the
tip voltage is $\approx 0.6$ and $\approx 2.2$ eV higher than the sample voltage.  These
correspond to the peaks that begin at 1.21 and 3.60 V in Fig. \ref{fig:dIdV}(b).
The disagreement in onset energies is not surprising, because electron disk energies $\epsilon_n$ 
become increasingly sensitive to the true boundary potential  as they get nearer to the Fermi energy $\epsilon_{Fs}$.  
The correct confining potential $V(z)$ from DFT theory for Eq. \ref{eq:uSW} has a smooth increase, 
not the soft-wall step function.  This influences higher $n$ energies more than lower ones. 

The $dI/dV$ spectra of Fig. \ref{fig:dIdV}(a) increase in height as $V$ increases, unlike
the simple model of Fig. \ref{fig:dIdV}(b) which shows no increase. 
It is sensible to modify Eq. \ref{eq:dIdV} by including the probability
$f_n^{\rm out}$ that state $n$ lies above the film.  The tunneling electrons that start from higher levels
(closer to $\epsilon_F$) have significantly larger probabilities of contacting the tip.
The modified equation is
\begin{equation}
\frac{dI}{dV} \propto \sum_{n=n_{\rm min}}^{n_{\rm max}} f_n^{\rm out} e^{-\gamma(eV+\epsilon_F-\epsilon_n)}. 
\label{eq:dIdV2}
\end{equation}
This will modify Fig. \ref{fig:dIdV}(b) by decreasing the height of the upper peak by a factor
$f_8^{\rm out}/f_7^{\rm out}=1.54$, worsening slightly the correspondence with the data in Fig. \ref{fig:dIdV}.
A more correct tunneling matrix element $M_{k_s,j_t}$ is needed, but is too complicated for inclusion
in a simple model.

\section{Summary}
\label{sec:summary}

The PBC and HW models have been used in literature because correct DFT calculations are very time-consuming for
films of more than a very few layers.  The SW model is more realistic, and allows approximate treatment 
of film spectroscopy.  Comparing SW theory with experiment in Fig. \ref{fig:dIdV}, the appearence of
two peaks is correctly explained, but the details are not closely aligned. 
Nevertheless, film free-electron models are still a useful substitute for full DFT calculations,
and, if carefully interpreted, provide insight into properties of films of simple metals.

\begin{appendix}

\section{Details of Fermi Disks }
\label{sec:appendix}

\subsection{Summing Occupancies}

Let $i=(k,j,\sigma)$ (where $k$ is a 3-vector in bulk crystals and 2-vector in films, $j$ an integer, and $\sigma$ the spin)
label single-electron
state properties.  A property $f_i$  (such as single-electron energy $\epsilon_i$) can be summed over states.
If the sample dimensions are large compared to the primitive unit cell, $k$-states are closely spaced, and
the $k$ part of the sum can be written as an integral:
\begin{equation}
\sum_i^{1d}f_i=\frac{L}{2\pi}\sum_{j,\sigma} \int_{\rm BZ}dk f_i
\label{eq:1}
\end{equation}
\begin{equation}
\sum_i^{2d}f_i=\frac{A}{(2\pi)^2}\sum_{j,\sigma} \int_{\rm BZ}d^2 k f_i
\label{eq:2}
\end{equation}
\begin{equation}
\sum_i^{3d}f_i=\frac{\Omega}{(2\pi)^3}\sum_{j,\sigma} \int_{\rm BZ}d^3 k f_i
\label{eq:3}
\end{equation}
where $L,A,\Omega$ are the length, area, or volume of the corresponding sample.

We need to count the number $N$ and number density $\rho_{\rm film}=N/(AH)$ of occupied free-electron states of a film
of macroscopic area $A$ and finite thickness $H=N_L h$.  The property $f_i$ is the average occupancy $N_{\rm{occ},i}$,
and the label $i$ is $(\vec{k}_\parallel,n,\sigma)$,where $n$ is the disk number.  The two spin states 
$\sigma=\pm 1$ of electrons have
the same average occupancy.  Brillouin-zone boundaries are irrelevant; $k$-vectors and free-electron energies
are continuous across these boundaries.  All three boundary conditions have free-electron energies 
\begin{equation}
\epsilon(\vec{k},n)=\frac{\hbar^2 k_\parallel^2}{2m}+\epsilon_n.
\label{eq:}
\end{equation}
In all cases, $\epsilon_n=\hbar^2 k_n^2/2m$.  In the PBC case,
$k_n=2\pi n/H$, with $n$ an integer, negative, 0, or positive.
In the HW case, $k_n=\pi n/H$, with $n$ a positive integer.
In the SW case, $\epsilon_n$
is not {\it a priori} given; it has is determined by solving a quantum well Schr{\"o}dinger equation.
The states lie in disks in $\vec{k}$-space, at distances $k_n$ in the $k_z$ direction.
The SW disks have $n>0$, and the $k_n$'s are not spaced periodically.

The number density is then
\begin{eqnarray}
&&\rho_{\rm film}=\frac{1}{AH}\sum_i f_i \nonumber \\
&=& \frac{2}{(2\pi)^2 H}\sum_{n} \int_{k<k_{Fn}} 
d^2 k N_{\rm{occ}}(\vec{k},n),
\label{eq:4}
\end{eqnarray}
where the 2 in the numerator accounts for the two spin states.
At $T=0$, the occupancy is 1 for all states below the Fermi energy, and zero otherwise.  
Each disk has the same Fermi energy $\epsilon_{F0}=\hbar^2 k_{F0}^2/2m$,
where $k_{F0}$ is the radius of the disk $n=0$ with $\epsilon_{n=0}=0$  
(the $n=0$ disk is only occupied in the PBC case).
The $n^{\rm th}$ disk has radius $k_{Fn}$ where $\hbar^2 k_{Fn}^2 /2m+\epsilon_n \equiv \epsilon_{F,{\rm film}}$.
The film electron density (assumed to be the same as in bulk) is then
\begin{equation}
\rho_{\rm film}=\frac{2}{(2\pi)^2 H}\sum_n^{\rm occ} \pi k_{Fn}^2.
\label{eq:5}
\end{equation}

\subsection{Periodic Boundary Conditions}
\label{sec:PBC}

For PBC,  $(k_{F0}^{\rm PBC})^2 = (k_{Fn}^{\rm PBC})^2 + k_n^2$, where
the quantized $k_z$ is $k_n=(2\pi n/H)^2$. 
The values of $k_{F0}^{\rm PBC}$ and $\epsilon_{F,N_L}^{\rm PBC}$ are close to $k_F$ and $\epsilon_F$
for the bulk free electron gas, but their true values need to be carefully determined.  Here is how it is done
for the 6-layer Al (111) film.  Since the Fermi energy of bulk Al is 11.65 eV and $(\hbar^2/2m)(2\pi/H)^2$ is 
0.7642 eV (using $H=6a/\sqrt{3}$ and the lattice constant $a=4.05 \AA$), it is sensible to guess that the maximum
$n$ of occupied disks is $n_{\rm max}^2 < 11.65/0.7642$, or that $n_{\rm max}<3.90$.
The electron density is given by
\begin{eqnarray}
&&(2\pi H)\rho_{\rm film}=\sum_{n=-n_{\rm max}}^{n_{\rm max}} 
\left[\left(k_{F0}^{\rm PBC}\right)^2-\left(\frac{2\pi n}{H}\right)^2 \right] \nonumber \\
 && \ \ \ \ \ \ \ \ \ \ =(2n_{\rm max}+1) \times \nonumber \\
 &&\left[ (k_{F0}^{\rm PBC})^2
 -\left(\frac{2\pi}{H}\right)^2 \frac{n_{\rm max}}{3}(n_{\rm max}+1)\right]. \nonumber \\
 \label{eq:6}
\end{eqnarray}
Bulk Al has a fcc structure with unit cell volume $a^3/4$ and 3 
valence electrons per atom ($12/a^3$ per unit volume).  Using the guess $n_{\rm max}=3$,
\begin{equation}
2\pi H\frac{12}{a^3}=7 \left[\left(k_{F0}^{\rm PBC}\right)^2-4\left(\frac{2\pi }{H}\right)^2 \right] .
\label{eq:7}
\end{equation}
Then solve for $k_{F0}^{\rm PBC}$ using $H=6a/\sqrt{3}$:
\begin{eqnarray}
(k_{F0}^{\rm PBC}a)_{n_{\rm max}=3}&=& \sqrt{ \left( \frac{144\pi}{7\sqrt{3}} + \frac{4\pi^2}{3}  \right)} = 7.104
\nonumber \\
(\epsilon_{F0}^{\rm PBC})_{n_{\rm max}=3}&=&11.72{\rm eV}.
\label{eq:8}
\end{eqnarray}
These are close to the free-electron parameters of bulk Al, so $n_{\rm max}=3$ was a good guess. 
The choice $n_{\rm max}=2$ gives a larger $\epsilon_{F0}^{\rm PBC}$ which yields an incorrect (higher) total energy.
Just to be sure, try again with $n_{\rm max}=4$, giving
\begin{equation}
(k_{F0}^{\rm PBC}a)_{n_{\rm max}=4}= \sqrt{ \left( \frac{144\pi}{9\sqrt{3}} + \frac{20\pi^2}{9}  \right)} = 7.138.
\label{eq:9}
\end{equation}
This is not allowed, because $k_4 a=8\pi a/H$ is 7.255, so the $n=\pm 4$ disks would have 
imaginary Fermi radii.

\subsection{Hard Wall Boundary Conditions}
\label{sec:HW}

Equation \ref{eq:5} still applies, and Eq. \ref{eq:6} becomes
\begin{eqnarray}
&&2\pi H \rho_{\rm film}=\sum_{n=1}^{n_{\rm max}}  \left[\left(k_{F0}^{\rm HW}\right)^2-\left(\frac{\pi n}{H}\right)^2 \right].
\nonumber \\
&=& n_{\rm max}\left[ \left(k_{F0}^{\rm HW}\right)^2 - \left(\frac{\pi}{H}\right)^2 \frac{n_{\rm max}+1}{6}
(2n_{\rm max}+1) \right]. \nonumber \\
\label{eq:10}
\end{eqnarray}
Soving for $k_{F0}^{\rm HW} a$,
\begin{equation}
(k_{F0}^{\rm HW} a)^2=\frac{144\pi}{n_{\rm max}\sqrt{3}} +\frac{\pi^2}{12} \frac{n_{\rm max}+1}{6}
(2n_{\rm max}+1)
\label{eq:}
\end{equation}
Following sec. \ref{sec:PBC}, it is sensible to guess that $n_{\rm max}=7$, which gives
\begin{eqnarray}
\left( k_{F0}^{\rm HW} a \right)_{n_{\rm max}=7}&=& \sqrt{\frac{144\pi}{7\sqrt{3}} + \frac{5\pi^2}{3}}=7.332
\nonumber \\
(\epsilon_{F0}^{\rm HW})_{n_{\rm max}=7}&=&12.49{\rm eV}.
\label{eq:12}
\end{eqnarray}
Now test the guess $n_{\rm max}=8$
\begin{equation}
\frac{12}{a^3}=\frac{1}{2\pi H} \left[8 \left(k_{F0}^{\rm HW}\right)^2-204\left(\frac{\pi }{H}\right)^2 \right].
\label{eq:13}
\end{equation}
Solving this gives
\begin{eqnarray}
\left( k_{F0}^{\rm HW} a \right)_{n_{\rm max}=8}&=& \sqrt{\frac{144\pi}{8\sqrt{3}} + \frac{17\pi^2}{8}}=7.323
\nonumber \\
(\epsilon_{F0}^{\rm HW})_{n_{\rm max}=8}&=&12.45{\rm eV}.
\label{eq:14}
\end{eqnarray}
This is a better guess.  The lower value of $\epsilon_{F0}^{\rm HW}$ yields a better (lower) total 
energy.  But it needs to be checked whether $n_{\rm max}=9$ works:
\begin{equation}
\frac{12}{a^3}=\frac{1}{2\pi H} \left[9 \left(k_{F0}^{\rm HW}\right)^2-285\left(\frac{\pi }{H}\right)^2 \right].
\label{eq:15}
\end{equation}
Solving this gives
\begin{equation}
\left( k_{F0}^{\rm HW} a \right)_{n_{\rm max}=9}= \sqrt{\frac{144\pi}{9\sqrt{3}} + \frac{95\pi^2}{36}}=7.421
\label{eq:16}
\end{equation}
This fails because $k_9 a=9\pi a/H$ is 8.162, so the $n= 9$ disk would have  
an imaginary Fermi radius.  The correct answer is $n_{\rm max}=8$, or 8 Fermi disks.

\subsection{Soft Wall Boundary Conditions}
\label{sec:SW}

We need to determine how many ($n_{\rm max}^{\rm SW}$) soft-wall quantum well states contribute occupied electrons
in the ground state.  Equation \ref{eq:10} is re-written as
\begin{equation}
\frac{\hbar^2}{2m}2\pi H \rho_{\rm film}=\sum_{n=1}^{n_{\rm max}} \left[ \epsilon_{F0}^{\rm SW}
-\epsilon_n^{\rm SW}\right].
\label{eq:17}
\end{equation}
This is a sum of states occupied in disks of radius $k_{Fn}^{\rm SW}=\sqrt{[(2m/\hbar^2)
(\epsilon_{F0}^{\rm SW}-\epsilon_n^{\rm SW})]}=\sqrt{2m\epsilon_n^{\rm SW}/\hbar^2}$, 
where the Fermi energy $\epsilon_{F0}^{\rm SW}$
is not yet known.  

Equation \ref{eq:17} can be rewritten as 
\begin{equation}
\frac{144\pi}{\sqrt{3}} \frac{\hbar^2}{2ma^2}= n_{\rm max} \epsilon_{F0}^{\rm SW}
-\sum_{n=1}^{n_{\rm max}}\epsilon_n^{\rm SW}.
\label{eq:18}
\end{equation}
Now guess a value for $n_{\rm max}$, solve for the corresponding $\epsilon_{F0}^{\rm SW}$,
and verify that the value of $\epsilon_{n_{\rm max}}^{\rm SW}$ is not greater than $\epsilon_{F0}^{\rm SW}$.
If not, then test by using larger guesses for $n_{\rm max}$.  A sensible guess is $n_{\rm max}=8$, as was found
in the HW case.  The sum of the first 8 quantum well energies in table I of the main text is 33.07 eV.  This yields
an SW Fermi energy of 11.72 eV, larger than the $8^{\rm th}$ quantum well energy.  Now test the case $n_{\rm max}=9$.
This gives a SW Fermi energy of 11.87 eV, smaller than the $9^{\rm th}$ quantum well energy, so 
$n_{\rm max}^{\rm SW}=8$ is correct, and $\epsilon_{F0}^{\rm SW}=11.72$ eV, accidentally the same as in the PBC
case, and very close to the bulk $\epsilon_F$ of Al, 11.65 eV.

\end{appendix}

\acknowledgements
I thank Mengkun Liu for discussions; M. Roesner, M. Y. Chou, and T.-C. Chiang for help with the literature;
and referees for helpful suggestions.

The author has no conflicts to disclose.

\bibliographystyle{unsrt}
\bibliography{film-free-electron-July}

\end{document}